\documentstyle[preprint,eqsecnum,prb,aps]{revtex}

\begin{document}

\title{Quadratic electronic response of a two-dimensional electron gas} 
\author{A. Bergara$^1$, J.M. Pitarke$^1$, and P. M. Echenique$^2$} 
\address{$^1$Materia Kondentsatuaren Fisika Saila, Zientzi
Fakultatea, Euskal Herriko  Unibertsitatea,\\ 644 Posta Kutxatila, 48080 Bilbo,
Basque Country, Spain\\ 
$^2$Materialen Fisika Saila, Kimika Fakultatea, Euskal Herriko
Unibertsitatea,\\  1072 Posta Kutxatila, 20080 Donostia, Basque Country, Spain}
\date\today
\maketitle

\begin{abstract}
The electronic response of a two-dimensional (2D) electron system represents a key quantity in
discussing one-electron properties of electrons in semiconductor heterojunctions, on the
surface of liquid helium and in copper-oxide planes of high-temperature superconductors. We
here report an evaluation of the wave-vector and frequency dependent dynamical quadratic
density-response function of a 2D electron gas (2DEG), within a self-consistent field
approximation. We use this result to find the $Z_1^3$ correction to the stopping power of a 2DEG
for charged particles moving at a fixed distance from the plane of the 2D sheet, $Z_1$ being the
projectile charge. We reproduce, in the high-density limit, previous full nonlinear
calculations of the stopping power of a 2DEG for slow antiprotons, and we go
further to calculate the $Z_1^3$ correction to the stopping power of a 2DEG for a wide range of
projectile velocities. Our results indicate that linear response calculations are, for all
projectile velocities, less reliable in two dimensions than in three dimensions.
\end{abstract}



\section{Introduction}

Since the pioneering work of Bohm and Pines\cite{Bohm}, the conduction electrons in a
metal have been described as a three-dimensional (3D) gas of electrons in a neutralizing uniform
positive charge. The dynamical linear density-response
function of a 3D electron gas (3DEG) was evaluated by Lindhard\cite{Lindhard} in the so-called
random-phase approximation (RPA), in which each electron is assumed to move in the external field
plus the induced field of all electrons. The wave-vector and frequency dependent dynamical quadratic
density-response function of a 3DEG has also been evaluated\cite{Cenni}, by going beyond linear
response theory. The knowledge of this quantity has been proved to be of great importance in
discussing the properties of electrons in a variety of 3D systems\cite{Richardson}, and, in
particular, in explaining the experimentally observed difference between the electronic energy
losses of protons and antiprotons moving through a solid\cite{Pitarke1,Pitarke2}.

The suggested existence of two-dimensional electron layers in metal-insulator-semiconductor
structures and on the surface of liquid helium led several years ago to a great activity in the
study of a two-dimensional electron gas\cite{Ando}, where the electrons are confined to a plane
and neutralized by an inert uniform rigid positive plane background. The 2D electron system has
also been considered in discussing the physics of new-class materials such as copper-oxide
planes of high-temperature superconductors\cite{Engel}. It has been found that electrons confined to
a 2D layer show sometimes interesting properties not shared by a 3D electron system. For instance,
for a 2D metal the plasma frequency goes to zero in the long-wavelength limit, in contrast to the 3D
situation\cite{Ritchie}. Stern\cite{Stern} evaluated the dynamical linear density-response function
of a 2DEG in the RPA, and calculated the plasmon dispersion and the asymptotic screened Coulomb
potential. Also, much effort has
gone into studying the ground state energy\cite{Rajagopal} and the excitation spectrum of a
2DEG\cite{Chaplik}. The stopping power for a fast particle moving parallel to a 2DEG was first
evaluated in the RPA, within linear response theory, by Horing {\it et al}\cite{Horing}, and the
effect on this quantity of finite temperature\cite{Bret}, local field corrections\cite{Wang1} and
recoil\cite{Bergara} has also been considered. Nonlinear calculations of the stopping power for
slow protons and antiprotons have been performed only very recently\cite{Perkus,Wang2} on the basis
of a scattering theory approach, the scattering cross sections being calculated for a statically
screened potential.

In this paper we present results for the dynamical quadratic electronic
density-response of a 2DEG to longitudinal external fields of arbitrary wave-vector and
frequency, which we evaluate on the same level of approximation as the RPA linear
density-response function of Stern\cite{Stern}. In order to illustrate the usefulness of the
knowledge of this quantity, we 
consider, as an example, the
stopping power of a 2DEG, and we use the quadratic density-response function to evaluate the
$Z_1^3$ nonlinear correction to the stopping power of a 2DEG for particles of charge $Z_1e$ moving
at a fixed distance from the 2D plasma.

\section{Dynamical electronic response}
\label{sec:level2}

We consider a uniform electron system of density $n_0$ at zero temperature. Linear and quadratic
density-response functions of this system, $\chi(x,x')$ and $Y(x,x',x'')$, with $x=({\bf r},t)$, may
be introduced in connection with the response to the presence of a time-varying external influence,
after an expansion of the induced electron density in powers of the external potential $V(x)$.
According to time-dependent perturbation theory, the induced electron density is given, up to second
order in $V(x)$, by (we use atomic units throughout, i.e.,
$e^2=\hbar=m_e=1$)
\begin{equation}
n^{ind}(x)=\int{\rm d}{x}'\chi(x,x')V(x')
+\int{\rm d}{x}'\int{\rm d}{x}''Y(x,x',x'')V(x')V(x''),
\end{equation}
where
\begin{equation} 
\chi(x,x')= 
-{\rm i}\,\Theta(t-t')<\Psi_0|\left[\tilde\rho_H(x),\tilde\rho_H(x')\right]|\Psi_0>
\label{1}
\end{equation}
and
\begin{eqnarray} 
Y(x,x',x'')=&&\Theta(t-t')\Theta(t'-t'')<\Psi_0|\left[\left[\tilde\rho_H(x),\tilde\rho_H(x')\right]
,\tilde\rho_H(x'')\right]|\Psi_0>/2\cr 
+&&\Theta(t-t'')\Theta(t''-t')<\Psi_0|\left[\left[\tilde\rho_H(x),\tilde\rho_H(x'')
\right],\tilde\rho_H(x')\right]|\Psi_0>/2
.
\label{1}
\end{eqnarray}
Here $|\Psi_0>$ denotes the normalized ground state, $\tilde\rho_H$ is the
density fluctuation operator $\tilde\rho_H=\hat\rho_H-n_0$, where $\hat\rho_H$ is the exact
Heisenberg density operator in the unperturbed system, and $\Theta(x)$ is the Heaviside step
function accounting for causality.

In a self-consistent field or random-phase approximation (RPA), it is assumed that the electron
density induced by an external potential can be replaced by the electron density induced in a
non-interacting electron gas by the sum of the external potential, $V(x)$, and the potential created
by the induced electron density itself, $V^{ind}(x)$. Hence, in this approximation:
\begin{eqnarray}
n^{ind}(x)=&&\int{\rm d}{x}'\chi^0(x,x')\left[V(x')+V^{ind}(x')\right]\cr
+&&\int{\rm d}{x}'\int{\rm d}{x}''
Y^0(x,x',x'')\left[V(x')+V^{ind}(x')\right]\left[V(x'')+V^{ind}(x'')\right].
\end{eqnarray}
Here $\chi^0(x,x')$ and $Y^0(x,x',x'')$ are 'free-particle' linear and quadratic
density-response functions, and
\begin{equation}
V^{ind}(x)=\int{\rm d}{x}'v(x,x')n^{ind}(x'),
\end{equation}
where $v(x,x')$ represents the instantaneous Coulomb interaction. Introducing Eq. (2.5) into Eq.
(2.4) and keeping in Eq. (2.4) only terms up to second order in $V(x)$, one finds the following
integral equations for RPA linear and quadratic density-response functions:
\begin{equation}
\chi(x_1,x_2)=\chi^0(x_1,x_2)
+\int{\rm d}{x_1'}\int{\rm d}{x_2'}\chi^0(x_1,x_1')v(x_1',x_2')\chi(x_2',x_2)
\end{equation}
and
\begin{eqnarray}
Y(x_1,x_2,x_3)=&&\int{\rm d}{x_2'}\int{\rm d}{x_3'}Y^0(x_1,x_2',x_3')K(x_2',x_2)K(x_3',x_3)\cr
+&&\int{\rm d}{x_1'}\int{\rm d}{x_1''}\chi^0(x_1,x_1')v(x_1',x_1'')Y(x_1'',x_2,x_3),
\end{eqnarray}
where $K(x,x')$ is the so-called inverse dielectric function:
\begin{equation}
K(x,x')=\delta(x-x')+\int{\rm d}{x''}v(x,x'')\chi(x'',x').
\end{equation}

The integral equations (2.6) and (2.7) have, within many-body perturbation theory, the simple
diagrammatic interpretation shown in Fig. 1, where empty and full bubbles represent non-interacting
and interacting linear density-response functions,
$\chi^0(x_1,x_2)$ and $\chi(x_1,x_2)$, respectively. Similarly, empty and full triangles represent
non-interacting and interacting quadratic density-response functions, $Y^0(x_1,x_2,x_3)$ and
$Y(x_1,x_2,x_3)$, respectively, and dashed lines represent the Coulomb interaction, $v(x,x')$. Thus,
RPA linear and quadratic density-response functions are represented diagrammatically by summing over
the infinite set of diagrams containing one (see Fig. 1a) and three (see Fig. 1b) strings of empty
bubbles, respectively.   

In the case of a homogeneous 2DEG, the electrons are free to move in two spatial dimensions, having
their motion constrained in the third dimension. Thus, assuming time invariance, we define the
Fourier transforms
\begin{equation}
\chi_q=\int{\rm d}^2{\bf r}_1\int{\rm d}t_1{\rm e}^{-{\rm i}\left[{\bf q}\cdot({\bf r}_1-{\bf
r}_2)-\omega(t_1-t_2)\right]}\chi({\bf r}_1,t_1;{\bf r}_2,t_2)
\end{equation}
and
\begin{eqnarray}
Y_{q_1,q_2}=\int{\rm d}^2{\bf r}_1\int{\rm d}t_1
\int{\rm d}^2{\bf r}_2\int{\rm d}t_2\,&&{\rm e}^{-{\rm i}\left[{\bf q}_1\cdot({\bf r}_1-{\bf
r}_2)-\omega_1(t_1-t_2)\right]}
{\rm e}^{-{\rm i}\left[({\bf q}_1+{\bf q}_2)\cdot({\bf r}_2-{\bf
r}_3)-(\omega_1+\omega_2)(t_2-t_3)\right]}\cr
&&\times Y({\bf r}_1,t_1;{\bf r}_2,t_2;{\bf r}_3,t_3),
\end{eqnarray}
where ${\bf r}_1$, ${\bf r}_2$ and ${\bf r}_3$ represent two-dimensional position-vectors in the 2D
plane, and $q$ is the trimomentum $q=({\bf q},q^0)$. Hence, within the RPA we find:
\begin{equation}
\chi_{q}=\chi^0_{q}+\chi^0_{q}v_{q}\chi_{q}
\end{equation}
and
\begin{equation}
Y_{q,-q_1}=Y_{q,-q_1}^0K_{q_1}K_{q-q_1}+\chi^0_{q}v_{q}Y_{q,-q_1},
\end{equation}
where
\begin{equation}
K_{q}=1+v_{q}\chi_{q}
\end{equation}
and $v_q=2\pi/|{\bf q}|$.  

For a non-interacting Fermi gas, the ground state is
obtained by filling all the plane wave states inside the Fermi sphere of radius $q_F=\sqrt 2/r_s$,
$r_s$ being the average interelectronic distance ($n_0^{-1}=\pi r_s^2$). As in the case of a
3DEG\cite{Zaremba0}, we find non-interacting linear and quadratic density-response functions to
be\cite{note1}
\begin{equation}
\chi_q^0=2\int{{\rm d}^2{\bf k}\over(2\pi)^2}
\left[{n_{\bf k}(1-n_{{\bf k}+{\bf q}})\over q^0-(\omega_{{\bf k}+{\bf q}}-\omega_{\bf k})+{\rm
i}\eta}+{(1-n_{{\bf k}})n_{{\bf k}+{\bf q}}\over -q^0-(\omega_{\bf k}-\omega_{{\bf k}+{\bf q}})-{\rm
i}\eta}\right]
\label{34}
\end{equation}
and
\begin{eqnarray}
Y_{q,-q_1}^0=-\int{{\rm d}^2{\bf k}\over (2\pi)^2}\bigl[
&&{n_{\bf k}(1-n_{{\bf k}+{\bf q}})(1-n_{{\bf k}+{\bf q}_1})\over(q^0+\omega_{\bf k}-\omega_{{\bf k}+{\bf q}}
+{\rm i}\eta)(q_1^0+\omega_{\bf k}-\omega_{{\bf k}+{\bf q}_1}+{\rm i}\eta)}\cr
&-&{(1-n_{\bf k})n_{{\bf k}+{\bf q}}n_{{\bf k}+{\bf q}_1}\over
(q^0+\omega_{\bf k}-\omega_{{\bf k}+{\bf q}}+{\rm i}\eta)
(q_1^0+\omega_{\bf k}-\omega_{{\bf k}+{\bf q}_1}+{\rm i}\eta)}\cr
&+&{n_{{\bf k}+{\bf q}}(1-n_{\bf k})(1-n_{{\bf k}+{\bf q}_1})\over
(-q^0+\omega_{{\bf k}+{\bf q}}-\omega_{\bf k}-{\rm i}\eta)
(-q^0+q_1^0+\omega_{{\bf k}+{\bf q}}-\omega_{{\bf k}+{\bf q}_1}-{\rm i}\eta)}\cr
&-&{(1-n_{{\bf k}+{\bf q}})n_{\bf k} n_{{\bf k}+{\bf q}_1}\over
(-q^0+\omega_{{\bf k}+{\bf q}}-\omega_{\bf k}-{\rm i}\eta)
(-q^0+q_1^0+\omega_{{\bf k}+{\bf q}}-\omega_{{\bf k}+{\bf q}_1}-{\rm i}\eta)}\cr
&+&{n_{{\bf k}+{\bf q}_1}(1-n_{\bf k})(1-n_{{\bf k}+{\bf q}})\over
(-q_1^0+\omega_{{\bf k}+{\bf q}_1}-\omega_{\bf k}-{\rm i}\eta)
(q^0-q_1^0+\omega_{{\bf k}+{\bf q}_1}-\omega_{{\bf k}+{\bf q}}+{\rm i}\eta)}\cr
&-&{(1-n_{{\bf k}+{\bf q}_1})n_{\bf k} n_{{\bf k}+{\bf q}}\over
(-q_1^0+\omega_{{\bf k}+{\bf q}_1}-\omega_{\bf k}-{\rm i}\eta)
(q^0-q_1^0+\omega_{{\bf k}+{\bf q}_1}-\omega_{{\bf k}+{\bf q}}+{\rm i}\eta)}\cr\cr
&+&(q_1\rightarrow q-q_1)
\bigr], 
\label{38} 
\end{eqnarray}
where $n_{\bf q}=\Theta(q_F-|{\bf q}|)$, $\omega_{\bf k}={\bf k}^2/2$, and $\eta$ is a positive
infinitesimal.

Analytical evaluation of Eq. (2.12) results in the non-interacting linear density-response
function of Stern\cite{Stern}. As for the non-interacting quadratic density-response function, we
first sum occupation numbers in Eq. (2.15) to find
\begin{eqnarray} 
Y_{q,-q_1}^0=-\int{{\rm d}^2{\bf k}\over(2\pi)^2} n_{\bf k} \bigl[&&{1\over
q^0+\omega_{\bf k}-\omega_{{\bf k}+{\bf q}}+{\rm i}\eta}\,{1\over
q_1^0+\omega_{\bf k}-\omega_{{\bf k}+{\bf q}_1}+{\rm i}\eta}\cr &+&{1\over
-q^0+\omega_{\bf k}-\omega_{{\bf k}+{\bf q}}-{\rm i}\eta}\,{1\over
-(q^0-q_1^0)+\omega_{\bf k}-\omega_{{\bf k}+{\bf q}-{\bf q}_1}-{\rm i}\eta}\cr &+&{1\over
-q_1^0+\omega_{\bf k}-\omega_{{\bf k}+{\bf q}_1}-{\rm i}\eta}\,{1\over
(q^0-q_1^0)+\omega_{\bf k}-\omega_{{\bf k}-({\bf q}-{\bf q}_1)}+{\rm i}\eta}\cr\cr
&+&(q_1\rightarrow q-q_1)
\bigr].
\label{39}
\end{eqnarray}
For the real part, we find\cite{note2}:  
\begin{equation}
{\rm Re}\left[
Y_{q,-q_1}^0\right]=-\left[(I_{q,q_1}+I_{q,q_1}')+(I_{-q,-q+q_1}+I_{-q,-q+q_1}')+
(I_{-q_1,q-q_1}-I_{-q_1,q-q_1}')\right],\label{9}
\end{equation}
where
\begin{eqnarray}
I_{q,q_1}&=&{1\over 2\pi|{\bf q}||{\bf q}_1|\sin\chi}
\{\arctan\left[{A\sin\chi\over A_1-A\cos\chi}\right]
-{\rm sgn}A\arctan\left[{\sin\chi\sqrt{A^2-q^2_F}\over A_1-A\cos
\chi}\right]\Theta (A^2-q^2_F)\cr
&+& 
\arctan\left[{A_1\sin\chi\over A-A_1\cos\chi}\right]
-{\rm sgn}A_1\arctan\left[{\sin\chi\sqrt{A_1^2-q^2_F}\over A-A_1\cos
\chi}\right]\Theta (A^2-q^2_F)\}\label{10}
\end{eqnarray}
and
\begin{eqnarray}
I_{q,q_1}'=-\Theta (q_F^2\sin^2 \chi-G){1\over 4|{\bf q}||{\bf q}_1|\sin\chi}.\label{10}
\end{eqnarray}
Here $A=q^0/|{\bf q}|-|{\bf q}|/2$,
$A_1=q_1^0/|{\bf q}_1|-|{\bf q}_1|/2$, and $G=\sqrt{A^2-2AA_1\cos\chi+A_1^2}$, $\chi$ being the
angle  between ${\bf q}$ and ${\bf q}_1$.

As for the imaginary part of $Y_{q,-q_1}^0$ we define\cite{Pitarke1,Pitarke2} the function
$H_{q,q_1}$, which can be represented in terms of a sum over hole and particle states and gives the
second order contribution to the so-called absorption probability, as demonstrated in
Ref.\onlinecite{Pitarke2}. We find, 
\begin{equation}  
{\rm Im}\left[Y_{q,-q_1}^0\right]=H_{q,q_1}+H_{q_1,q}+H_{q-q_1,-q_1},\label{14}
\end{equation}
where\cite{note2}
\begin{equation}
H_{q,q_1}={1\over 2}\left[f_{q,q_1}-f_{-q,-q+q_1}+(q_1\rightarrow q-q_1)\right]\label{17}
\end{equation}
and
\begin{equation}
f_{q,q_1}=\Theta(q_F^2-A^2){1\over 2\pi|{\bf q}||{\bf q}_1|\sin\chi}
\ln \left | {A_1-A\cos\chi+\sin\chi \sqrt{q_F^2-A^2}\over
A_1-A\cos\chi-\sin\chi \sqrt{q_F^2-A^2}}\right |.\label{18}
\end{equation}

In particular, at low frequencies an expansion of $H_{q,q_1}$ in powers of the frequency $q^0$
gives, after retaining only the first-order terms,
\begin{equation}
H_{q,q_1}^L=\Theta (4q^2_F-|{\bf q}|^2){4(|{\bf q}|\cos\chi-|{\bf q}_1|)\over\pi|{\bf q}||{\bf
q}_1|\left[|{\bf q}-{\bf q}_1|^2-4q^2_F\sin^2\chi\right]\sqrt{4q^2_F-|{\bf q}|^2}}\,q^0.\label{20}
\end{equation}
In the static limit ($q^0\to 0$) both first and second order contributions to the
absorption probability, ${\rm Im}\chi_q$ and
$H_{q,q_1}$, are proportional to the frequency $q^0$, as in the case of a 3DEG.

\section{Electronic stopping power}

We consider an ion of charge $Z_1$ moving with constant velocity ${\bf v}$ at a fixed
distance $h$ from a 2DEG of density $n_0$. The Coulomb potential of this moving particle has the
form
\begin{equation}
V({\bf r},z;t)=Z_1|{\bf r}-{\bf v}t+(z-h){\bf\hat k}|^{-1},
\end{equation}
where ${\bf r}$ represents, as in Eqs. (2.9) and (2.10), a two-dimensional position-vector in the 2D
plane, and $z$ denotes the coordinate normal to the 2DEG which we consider to be located at
$z=0$. Hence, in order to obtain the induced potential, we introduce this time-varying external
potential into Eq. (2.1), and Eq. (2.1) into Eq. (2.5). We note that density-response functions of a
2DEG have their $z$ arguments localized to the 2D plane by positional $\delta$ functions, we Fourier
transform, and find, up to second order in
$Z_1$:
\begin{eqnarray} 
V^{ind}({\bf r},z;t)=&&Z_1\int{{\rm d}^2{\bf
q}\over(2\pi)^2}{\rm e}^{{\rm i}({\bf q}\cdot{\bf r}-\omega t)-|{\bf q}|(|z|+h)}v_{{\bf q}}
\chi_{{\bf q},\omega}v_{\bf q}\cr 
+&&Z_1^2\int{{\rm
d}^2{\bf q}\over(2\pi)^2}\int{{\rm d}^2{\bf q}_1\over(2\pi)^2}{\rm e}^{{\rm i}({\bf q}\cdot{\bf r}-\omega t)-\left[|{\bf q}||z|+(|{\bf
q}_1|+|{\bf q}-{\bf q}_1|)h\right]}v_{{\bf q}}Y_{{\bf q},\omega;-{\bf
q}_1,-\omega_1}v_{{\bf q}_1}v_{{\bf q}-{\bf q}_1}, 
\label{60}
\end{eqnarray}
where $\omega={\bf q}\cdot{\bf v}$ and $\omega_1={\bf q}_1\cdot{\bf v}$. 

The stopping power of the 2DEG is simply the retarding force that the polarization charge
distribution in the vicinity of the projectile exerts on the projectile itself\cite{Echenique1}, and
is given by
\begin{equation} 
S={Z_1\over v}\int{\rm d}^2{\bf r}\int{\rm d}z\delta({\bf r}-{\bf v}t)\delta(z-h){\bf
\nabla}V^{ind}({\bf r},z;t)\cdot{\bf v}.\label{59}  
\end{equation}
Substituting Eq. (3.2) into Eq. (3.3), we have\cite{note3}
\begin{eqnarray} 
S=-&&{Z_1^2\over v}\int{{\rm d}^2{\bf
q}\over(2\pi)^2}\omega{\rm e}^{-2|{\bf q}|h}v_{{\bf q}}
{\rm Im}\left[\chi_{{\bf q},\omega}\right]v_{\bf q}\cr 
-&&{Z_1^3\over v}\int{{\rm
d}^2{\bf q}\over(2\pi)^2}\omega\int{{\rm d}^2{\bf q}_1\over(2\pi)^2}{\rm e}^{-(|{\bf
q}|+|{\bf q}_1|+|{\bf q}-{\bf q}_1|)h}v_{{\bf q}}{\rm Im}\left[Y_{{\bf
q},\omega;-{\bf q}_1,-\omega_1}\right]v_{{\bf q}_1}v_{{\bf q}-{\bf q}_1}. 
\label{60}
\end{eqnarray}

In the RPA, $\chi_q$ and $Y_{q,-q_1}$ are obtained from Eqs. (2.11) and (2.12) or, equivalently,
from
\begin{equation}
v_q\chi_q=K_q-1
\end{equation}
and
\begin{equation}
Y_{q,-q_1}=K_qY_{q,-q_1}^0K_{q_1}K_{q-q_1},
\end{equation}
where
\begin{equation}
K_q=\left(1-\chi_q^0v_q\right)^{-1}.
\end{equation}

The linear contribution to the stopping power of Eq. (3.4), which is proportional to $Z_1^2$, was
evaluated in the RPA by Horing {\it et al}\cite{Horing} at $T=0$, and similar calculations were
presented by Bret and Deutsch\cite{Bret} at finite temperature. On the other hand, the
quadratic contribution to the stopping power of Eq. (3.4) is, in the RPA and for a geometry with the
ion-beam in-plane within the 2D electron layer ($h=0$), equivalent to the result derived in
Ref.\onlinecite{Pitarke2}, within many-body perturbation theory, as the energy loss per unit path
length of the projectile, the integration space being now in two dimensions instead of three
dimensions. This contribution to the stopping power, which is proportional to
$Z_1^3$, discriminates between the energy loss of a proton and that of an antiproton, and appears as
a consequence of losses to one- and two-step electronic excitations generated by both linearly and
quadratically screened ion potentials, as discussed in Ref.\onlinecite{Pitarke2}.

In the case of slow intruders ($v\to 0$), only the low-frequency form of the response enters in the
evaluation of Eq. (3.4). At zero frequencies both linear and quadratic density-response functions
are real, ${\rm Im}\left[\chi^0_q\right]$ and
$H_{q,q_1}$ being at low frequencies proportional to the frequency $q^0$. Thus, retaining only the
first-order terms in the frequencies,  both
$Z_1^2$ and $Z_1^3$ contributions to the stopping power are found to be proportional to the velocity
of the projectile, as in a 3DEG. For the RPA quadratic contribution to the stopping power we find,
after insertion of Eq. (2.23) into Eq. (3.4), the following result:
\begin{equation}
\left[S^L\right]^{(3)}=8 v Z_1^3 \int_0^\infty{{\rm d}|{\bf q}|\over\sqrt{4q^2_F-|{\bf
q}|^2}}\int_0^\infty{\rm d}|{\bf q}_1|\int^{\pi}_0
{\rm d}\chi{\rm e}^{-(|{\bf
q}|+|{\bf q}_1|+|{\bf q}-{\bf q}_1|)h}{f^L_1+f^L_2\over|{\bf q}-{\bf q}_1|},\label{21}
\end{equation}
where
\begin{equation}
f^L_{1}=\Theta (2q_F-|{\bf q}|)K_{{\bf q},0}^{2} K_{{\bf q}_1,0} 
K_{{\bf q}-{\bf
q}_1,0} Y_{{\bf q},0;{\bf q}_1,0}^0\label{22}
\end{equation}
and 
\begin{equation}
f^L_{2}=-\Theta (2q_F-|{\bf q}|){{ |{\bf q}| \left(|{\bf q}|\cos \chi- |{\bf
q}_1|\right)}\over{\pi|{\bf q}_1|
\left(|{\bf q}-{\bf q}_1|^2-4q^2_F\sin^2\chi\right)}}K_{{\bf q},0}K_{{\bf q}_1,0} 
K_{{\bf q}-{\bf q}_1,0}.\label{23}
\end{equation}

\section{Results}

In the low-velocity limit ($v\to 0$), the stopping power can be evaluated to all orders in
$Z_1$, on the basis of a free-electron picture, with the additional assumption of independent,
individual, elastic electron scattering, and it is easily found to be, in both two and three
dimensions, proportional to the velocity of the projectile\cite{Nagy}. By using
density-functional theory (DFT)\cite{Kohn} to calculate the
self-consistent potential generated by a static charge submerged in a 3DEG, Echenique {\it
et al}\cite{Echenique2} evaluated the nonlinear stopping power of a 3DEG for slow ions. Nonlinear
calculations for an in-plane projectile in 2D have been performed recently by using a nonlinearly
screened scattering potential based on the Sj\"olander-Stott theory\cite{Perkus}, with the use of a
nonlinear version of the linearized Fermi-Thomas potential\cite{Wang2} in which the screening
constant is determined from the Friedel sum rule, and on the basis of a self-consistent
potential\cite{Zaremba} obtained within DFT as in Ref.\onlinecite{Echenique2} for the 3D case.

The nonlinear calculations presented in Ref.\onlinecite{Wang2} for the energy-loss
of slow protons and antiprotons are very close to the calculations of Ref.\onlinecite{Zaremba} when
exchange and correlation (XC) contributions to the DFT scattering potential are excluded. These
calculations are correct to all orders in $Z_1$, and they represent, therefore, a good check for our
quadratic response calculations, which should be exact in the high-density and/or low $Z_1$ limits.
Nevertheless, the approaches of Refs.\onlinecite{Perkus,Wang2} and
\onlinecite{Zaremba} have the limitation of being restricted to very low ion velocities ($v<<v_F$,
$v_F$ being the Fermi velocity), while our nonlinear $Z_1^3$ corrections are
valid for arbitrary non-relativistic velocities. 

The low-velocity limit of the
quadratic stopping power, as obtained from Eq. (3.8) for an in-plane ($h=0$) moving charge and
divided by the velocity, is represented in Fig. 2 by a solid line, as a function of the
electron-density parameter $r_s$. Stars and squares represent the full nonlinear contribution to the
stopping power for antiprotons reported in Refs.\onlinecite{Wang2} and \onlinecite{Zaremba},
respectively, multiplied by a factor of
$-1$, showing an excellent agreement, in the high-density limit, with our $Z_1^3$ nonlinear
contribution. For $r_s\ge 2$, higher order corrections become important, and for $r_s>3$, $Z_1^2$
contributions (dashed line) are smaller than the $Z_1^3$ correction, indicating that the external
potential cannot be treated, for these electron densities, as a small perturbation. In the case of a
projectile moving at a given $h$ distance above the 2D plasma the external perturbation is obviously
diminished and, in particular, for $h=1/q_F$ the quadratic stopping power is smaller than the linear
one for all electron densities (see the inset of Fig. 2). The full nonlinear contribution to the
stopping power for protons reported in Refs.\onlinecite{Wang2} and \onlinecite{Zaremba} for $h=0$ is
also represented in Fig. 2 by circles and crosses, respectively, showing large differences with our
$Z_1^3$ calculations for all electron densities. These differences appear as a consequence of
perturbation theory failing to describe electronic states bound to the proton, which in 2D systems
can be supported by arbitrarily weak attractive potentials\cite{Economou}.

As the velocity of the projectile ($v$) and/or the distance from the 2D plasma ($h$) increase, the
ion potential becomes a relatively smaller perturbation and the $Z_1^3$ contribution to the stopping
power for antiprotons may, therefore, be expected to approximately describe the full nonlinear
contribution to the stopping power for arbitrary values of $v$ and $h$, as long as $r_s<2$, and also
for lower densities ($r_s\ge 2$) as the velocity and the $h$ distance increase. Substitution of the
full RPA linear and quadratic 2DEG density-response functions of Eqs. (3.5) and (3.6) into Eq. (3.4)
results in the quadratic stopping power plotted by a solid line in Fig. 3, as a function of the
impinging projectile velocity, for
$r_s=1$ and $h=0$, and also for
$r_s=1$ and $h=1/q_F$ (see the inset of Fig. 3). The dotted line represents the low-velocity
limit, as obtained from Eq. (3.8), and dashed lines represent linear RPA contributions to
the stopping power of Eq. (3.4).

The quadratic contribution to the stopping power
of a 2DEG presents properties not shared by the 3DEG. First, the range of validity of the linear
velocity dependence of this contribution to the stopping power (see Fig. 3) persists only up to
velocities much smaller than the Fermi velocity, in contrast to the 3D situation in which the
linear velocity dependence persists up to velocities approaching the Fermi velocity\cite{Pitarke2}.
Second, at velocities around the plasmon threshold velocity, for which the projectile has enough
energy  to excite a plasmon\cite{note0}, the ratio between  $Z_1^3$ and $Z_1^2$ contributions to the
stopping power increases, again in contrast to the 3D situation. Furthermore, we have found that the
increase of this ratio at the plasmon threshold velocity becomes dramatic as the electron density
decreases. Of course, the ratio between quadratic and linear contributions to the stopping power
decreases very rapidly with $h$ (see the inset of Fig. 3), but the relative, and for large $r_s$
dramatic, increase of this ratio at the plasmon threshold velocity persists for all values of
$h$ that are not much larger than $1/q_F$. For a 2DEG the group velocity of the plasmon wave
nearly coincides with the plasmon threshold velocity of the projectile (see Fig. 4); also, the
inclusion of short-range correlations, which are ignored in the RPA, is known to have a substantial
effect on the plasmon dispersion\cite{Gold}. We interpret the anomalous enhancement of the ratio
between quadratic and linear contributions to the stopping power at the plasmon threshold velocity
and small electron densities ($r_s>1$) as a result of neglecting, within the RPA, short-range
correlations between the electrons of a 2D system, since these correlations are non-negligible for
all values of $h$ as long as $r_s$ is not small.

\section{Conclusions}

We have presented an analytical evaluation of the wave-vector and frequency dependent
non-interacting quadratic density-response function of a 2DEG. We have used this
result to find, within a self-consistent field approximation, the $Z_1^3$ correction to the stopping
power of a 2DEG for charged recoiless particles moving at a fixed distance from the 2D plasma sheet.
We have reproduced, in the high-density limit, previous full nonlinear calculations of the stopping
power of a 2DEG for slow antiprotons, and we have gone further to calculate the $Z_1^3$ correction
to the stopping power for a wide range of projectile velocities. We have found that the $Z_1^3$
contribution to the stopping power of a 2DEG presents properties not shared by the 3DEG. On the
one hand, the range of validity of the linear velocity dependence of the $Z_1^3$ contribution to the
stopping power persists only up to velocities much smaller than the Fermi velocity, and, on
the other hand, an anomalous enhancement of the ratio between quadratic and linear
contributions to the stopping power at the plasmon threshold velocity is found, within the RPA, at
small electron densities ($r_s>1$). Also, our results indicate that linear response
calculations are, for all projectile velocities, less reliable in 2D than in 3D.

\acknowledgments

We acknowledge partial support by the University of the Basque
Country, the Basque Unibertsitate eta Ikerketa Saila, the Spanish Ministerio de Educaci\'on y
Cultura, and Iberdrola SA.

\begin{figure}
\caption{Diagrammatic interpretation of the RPA integral equations (2.6) and (2.7) for (a)
linear and (b) quadratic density-response functions, respectively. Electron-hole empty and full
bubbles represent non-interacting and interacting linear density-response functions, respectively.
Empty and full triangles represent non-interacting and interacting quadratic density-response
functions, respectively. The interacting RPA quadratic density-response function is obtained by
summing over the infinite set of diagrams that combine three strings of empty bubbles (two-electron
loops) through an empty three-electron loop. Dashed lines represent the electron-electron bare
Coulomb interaction.}
\end{figure}

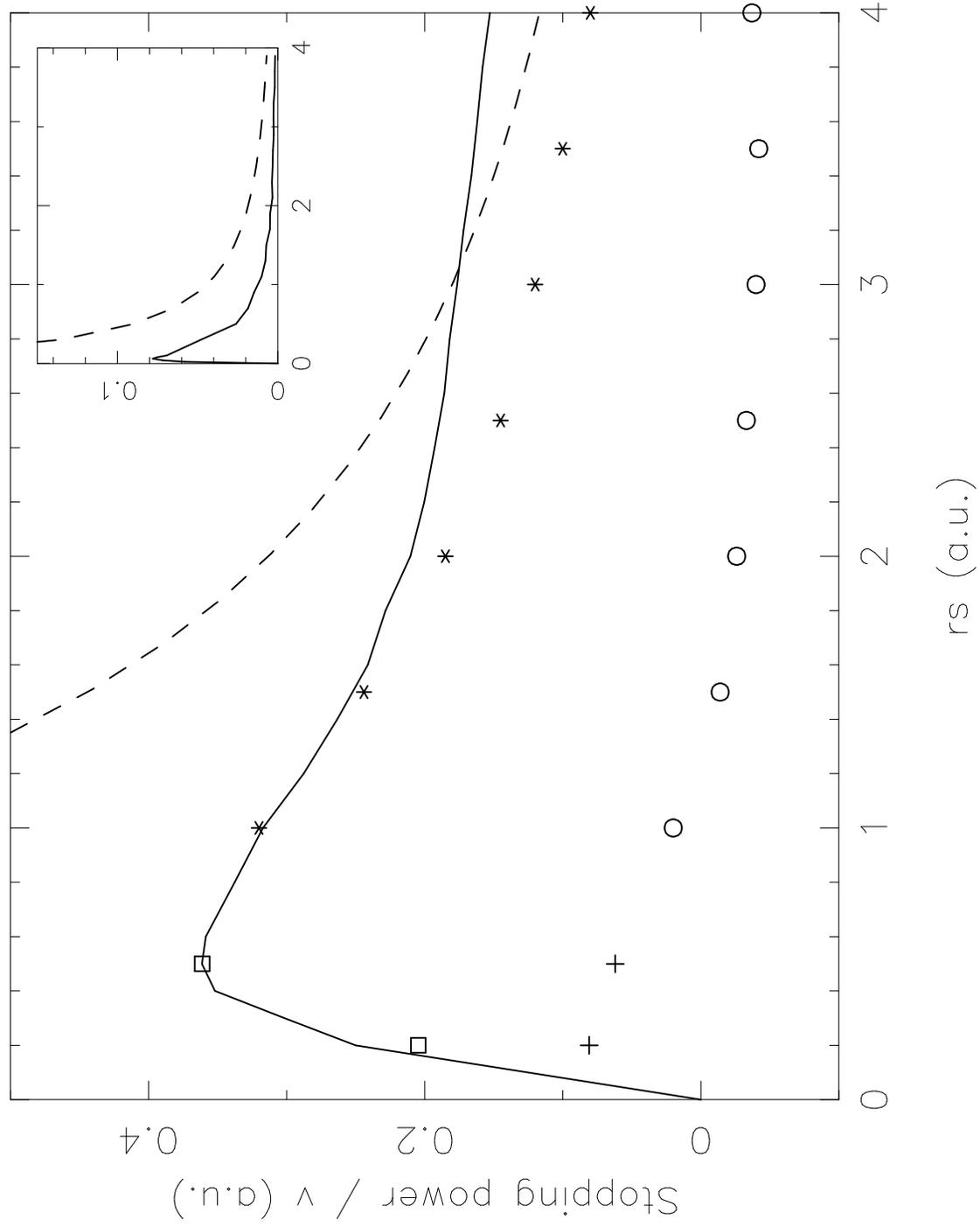
\begin{figure}
\caption{Low-velocity limit of the $Z_1^3$ stopping power, as obtained from Eq. (3.8) (solid line)
for $h=0$ and $Z_1=1$, divided by the velocity, as a function of $r_s$; the
corresponding $Z_1^2$ stopping power is represented by a dashed line. Full nonlinear contributions
to the stopping power for antiprotons [$Z_1=-1$] (stars and squares) and protons [$Z_1=1$] (circles
ans crosses) have been obtained by subtracting RPA linear calculations from the results of
Ref.\protect\onlinecite{Wang2}\protect$\,$(stars and circles) and Ref.
\protect\onlinecite{Zaremba}\protect$\,$[with XC contributions to the DFT
scattering potential excluded] (squares and crosses), and dividing by $Z_1$; thus, the negative
values at $r_s>1$ simply mean that the full nonlinear stopping power lies, in the case of protons,
below the linear results. The inset exhibits
$Z_1^2$ and $Z_1^3$ contributions to the stopping power (dashed and solid lines, respectively) for
$h=1/q_F$ and $Z_1=1$.}
\end{figure}

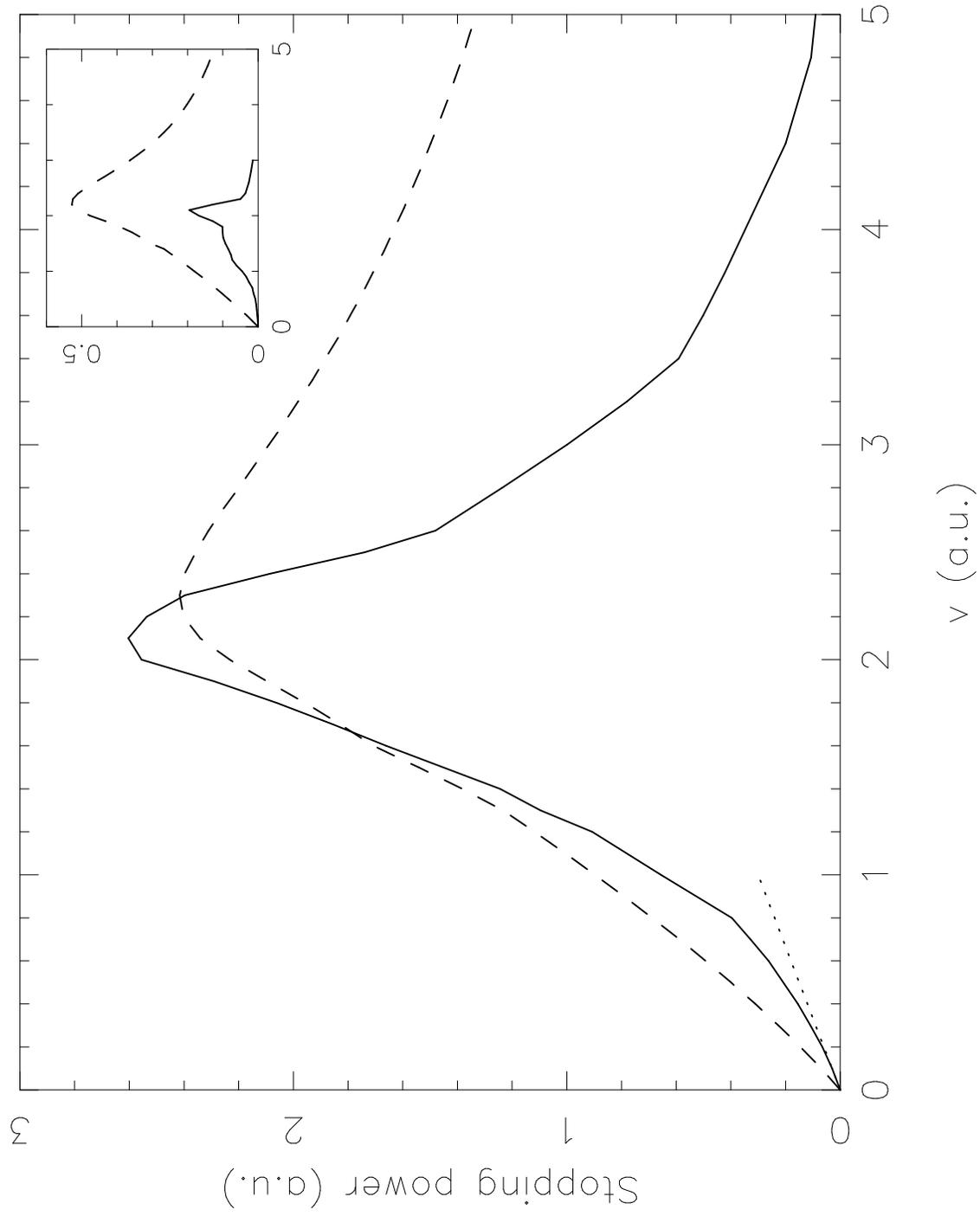
\begin{figure}
\caption{Full RPA $Z_1^3$ stopping power, as obtained from Eq. (3.4) (solid line) for $Z_1=1$,
$h=0$ and $r_s=1$, as a function of the velocity of the projectile. The corresponding $Z_1^2$
stopping power is represented by a dashed line, and the dotted line represents
the low-velocity limit of the $Z_1^3$ term. The inset shows the same results for
$h=1/q_F$.}
\end{figure}

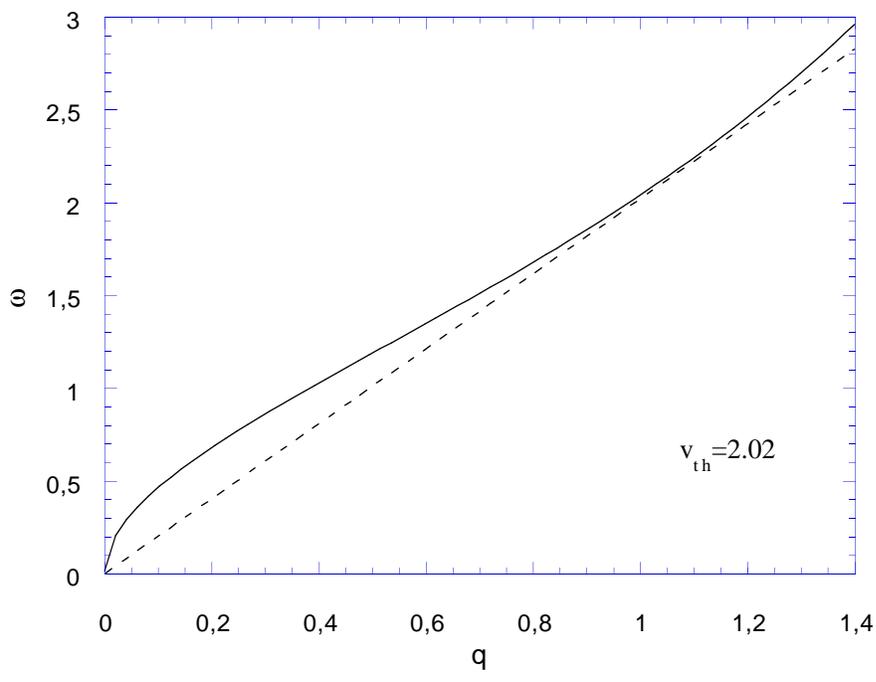
\begin{figure}
\caption{Two-dimensional RPA plasmon dispersion relation (solid line) and maximum energy
transfer ($\omega_{max}=qv$) at the plasmon threshold velocity (dotted line), as functions of the
wave number. The electron density parameter has been taken to be $r_s=1$, thus the plasmon threshold
velocity being $v_{th}=2.02v_0$ ($v_0$ is Bohr' velocity).}
\end{figure}  


\begin{references}
\bibitem{note2} $I_{q,q_1}'$, $I_{-q,-q+q_1}'$ and $I_{-q_1,q-q_1}'$ in Eq. (2.17) should be
multiplied by
${\rm sgn}(q^0){\rm sgn}(q_1^0)$, ${\rm sgn}(q^0){\rm sgn}(q^0-q_1^0)$ and
${\rm sgn}(q_1^0){\rm sgn}(q^0-q_1^0)$, respectively, to obtain the real part of the
corresponding time-ordered quadratic response function. Similarly, the right-hand side of Eq. (2.21)
should be multiplied by
${\rm sgn}(q^0)$ in order to obtain the imaginary part of the time-ordered quadratic response
function.
\bibitem{note3} The real parts of $\chi_q$ and $Y_{q,-q_1}$ give no contribution to the integrals
of Eq. (3.4), since ${\rm Re}\left[\chi_q\right]={\rm Re}\left[\chi_{-q}\right]$ and ${\rm
Re}\left[Y_{q,q_1}\right]={\rm Re}\left[Y_{-q,-q_1}\right]$.
\end{references}
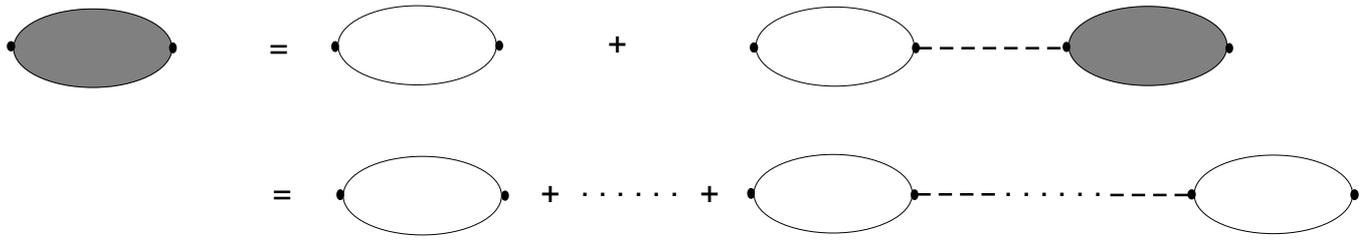
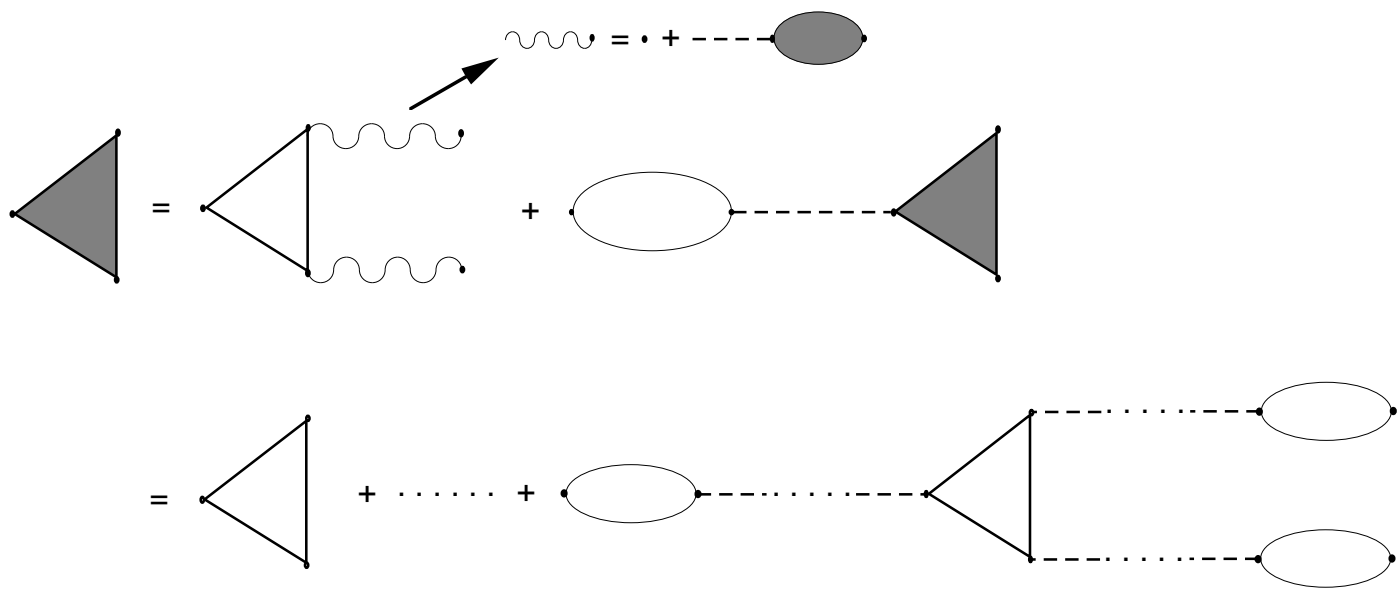
\end{document}